\documentclass[10pt,twocolumn,twoside]{IEEEtran}
\usepackage[inline]{enumitem}

\usepackage{xr}
\usepackage{multirow}

\usepackage{amssymb}

\usepackage{preamble}
\usepackage{mathtools}
\usepackage{cases}

\usepackage{booktabs}

\usepackage{subcaption}

\usepackage[style=ieee,citestyle=numeric-comp, natbib=false, sorting=none,
uniquename=false, hyperref=false, backend=bibtex, autolang=hyphen,
bibencoding=utf8, minnames=1,defernumbers=true, maxcitenames=2,
maxbibnames=6]{biblatex}

\AtEveryBibitem{\clearfield{note}}
\AtEveryBibitem{\clearfield{doi}}
\AtEveryBibitem{\clearfield{issn}}
\AtEveryBibitem{\clearfield{isbn}}

\title{Upper-Bounding the Regularization Constant for Convex Sparse Signal 
Reconstruction}
\author{\IEEEauthorblockA{Renliang Gu and Aleksandar Dogand\v{z}i{\'c}}
 \thanks{The authors are with the Department of Electrical and Computer 
   Engineering, Iowa State University, Ames, IA 50011 USA (e-mail:
    \texttt{\{\mbox{renliang},ald\}@iastate.edu}).
    This work was supported by the National Science Foundation under Grant 
  CCF-1421480.  }

}

\newtheorem{thm}{Theorem}
\newtheorem{rem}{Remark}
\newtheorem{exa}{Example}
\newtheorem{lem}{Lemma}
\newtheorem{cor}{Corollary}

\usepackage{todonotes}

\begin{document}
\maketitle

\begin{abstract}
  Consider reconstructing a signal $\bx$ by minimizing a weighted sum of a 
  convex differentiable \gls{NLL} (data-fidelity) term and a convex 
  regularization term that imposes a convex-set constraint on $\bx$ and 
  enforces its sparsity using $\bm{\ell}_1$-norm analysis regularization.  
  We compute upper bounds on the regularization tuning constant beyond 
  which the regularization term overwhelmingly dominates the \gls{NLL} term 
  so that the set of minimum points of the objective function does not 
  change.  Necessary and sufficient conditions for irrelevance of sparse 
  signal regularization and a condition for the existence of finite upper 
  bounds are established.  We formulate an optimization problem for finding 
  these bounds when the regularization term can be globally minimized by a 
  feasible $\bx$ and  also develop an \gls{ADMM} type method for their 
  computation.  Simulation examples show that the derived and empirical 
  bounds match.
\end{abstract}

 %We focus on the practically important scenario where the vector space 
 %orthogonal to the range of the sparsifying dictionary matrix intersects 
 %with the convex set that constrains the signal $\bx$. 

\glsresetall
\glsunset{CPU}

\section{Introduction}
\label{sec:intro}

Selection of the regularization tuning constant $u > 0$ in convex 
Tikhonov-type \cite{TikhonovArsenin} penalized \gls{NLL} minimization 
\begin{IEEEeqnarray}{rCl}
  \label{eq:penalizedNLL}
  f_u(\bx)=\cL(\bx) +u r(\bx)
\end{IEEEeqnarray}
is a challenging problem critical for obtaining accurate estimates of the 
signal $\bx$ \cite[Ch.~7]{Vogel2002}.  Too little regularization leads to 
unstable reconstructions with large noise and artifacts due to, for 
example, aliasing.  With too much regularization, the reconstructions  are 
too smooth and often degenerate to constant signals.  Finding bounds on the 
regularization constant $u$ or finding conditions for the irrelevance of 
signal regularization has received little attention.  In this paper, we 
determine upper bounds on $u$ beyond which the regularization term $r(\bx)$ 
overwhelmingly dominates the \gls{NLL} term $\cL(\bx)$ in 
\eqref{eq:penalizedNLL} so that the minima of the objective function 
$f_u(\bx)$ \emph{do not} change. For a linear measurement model with white 
Gaussian noise and $\ell_1$-norm regularization, a closed-form expression 
for such a bound is determined in \cite[eq.~(4)]{KimBoyd2007}; see also 
Example~\ref{ex:nonnegsig}.    The obtained bounds can be used to design 
continuation procedures \cite{HaleYin2008,Wright2009SpaRSA} that gradually 
decrease $u$ from a large starting point down to the desired value, which 
improves the numerical stability and convergence speed of the resulting 
minimization algorithm by taking advantage of the fact that penalized 
\gls{NLL} schemes converge faster for smoother problems with larger $u$ 
\cite{AllgowerGeorg2003}.  In some scenarios, users can monitor the 
reconstructions as $u$ decreases and terminate when the result is 
satisfactory.

Consider a convex \gls{NLL} $\cL(\bx)$ and a regularization term
\begin{IEEEeqnarray}{c}
  \label{eq:r}
  r(\bx) =  \mathbb{I}_C(\bx) + \norm{ \Psi^H \bx }_1   
\end{IEEEeqnarray}
that imposes a convex-set constraint on $\bx$, $\bx \in C \subseteq 
\mathamsbb{R}^p$, and sparsity of an appropriate linearly transformed 
$\bx$, where $\Psi \in \mathamsbb{C}^{p\times p'}$  is a known 
\emph{sparsifying dictionary} matrix.
Assume that the \gls{NLL} $\cL(\bx)$ is differentiable and lower bounded
within the closed convex set $C$, and satisfies
\begin{IEEEeqnarray}{c}
  \label{eq:Ccond}
  \dom \cL(\bx) \supseteq C
\end{IEEEeqnarray}
which ensures that  $\cL(\bx)$ is computable for all $\bx \in C$.
Define the convex sets of solutions to $\min_\bx f_u(\bx)$,
$\min_\bx r(\bx)$, and $\min_{\bx \in Q} \cL(\bx)$:\footnote{The use of 
  ``$\leq$'' in the definitions of $Q$ and $\cX^\diamond$
 in \eqref{eq:Q} and \eqref{eq:xdiamond} makes it easier to identify both 
 as convex sets.}
\begin{subequations}
\begin{IEEEeqnarray}{rCl}
  \label{eq:xustar}
  \cX_u
  &\df&
  \CBRbig{
    \bx \,\big|\,
    f_u(\bx)=\min_{\bchi}f_u(\bchi)
  }\\
  \label{eq:Q}
  Q &\df& \CBRbig{ \bx\mid  r(\bx) = \min_\bchi r(\bchi)  }
  \notag \\
  &=&
  \CBRbig{
    \bx\in C \,\big|\,  \norm{\Psi^H \bx}_1 \leq
    \min_{\bchi \in C} \norm{\Psi^H \bchi  }_1
  }
  \\
  \label{eq:xdiamond}
  \cX^\diamond
  &\df&
  \CBRbig{
    \bx\in Q \mid \cL(\bx) \leq \min_{\bchi \in Q} \cL(\bchi)
  }
  \neq\emptyset
\end{IEEEeqnarray}
\end{subequations}
where the existence of $\cX^\diamond$ is ensured by the assumption that 
$\cL(\bx)$ is lower bounded in $C$.

We review the notation: ``$^*$'', ``$^T$'', ``$^H$'', ``$^+$'', 
$\norm{\cdot}_p$, $\abs{\cdot}$, $\otimes$, ``$\succeq$'', ``$\preceq$'', 
$I_N$, $\bm{1}_{N \times 1}$, and $\bm{0}_{N\times1}$ denote complex 
conjugation, transpose, Hermitian transpose, Moore-Penrose matrix inverse, 
$\ell_p$-norm over the complex vector space $\mathamsbb{C}^N$ defined by 
$\|\bz\|_p^p = \sum_{i=1}^N |z_i|^p$ for $\bz = (z_i) \in \mathamsbb{C}^N$, 
absolute value, Kronecker product, elementwise versions of ``$\geq$'' and 
``$\leq$'', the identity matrix of size $N$ and the $N \times 1$ vectors of 
ones and zeros, respectively (replaced by $I,\bm{1}$, and $\bm{0}$ when the 
dimensions can be inferred).  $\mathbb{I}_C(\ba) = \ccases{ 0, & \ba \in C 
\\ +\infty, & \text{otherwise} }$,
$\proj{C}{\ba} = \arg \min_{\bx \in C} \norm{\bx-\ba}_2^2$,
and $\exp_\circ \ba$
denote the indicator function, projection onto $C$, and the elementwise 
exponential function: $\SBRs{\exp_\circ \ba }_i= \exp a_i$.

  Denote by $\mathcal{N}(A)$ and $\mathcal{R}(A)$ the null space and range 
  (column space)  of a matrix $A$. These vector spaces are real or complex 
  depending on whether $A$ is a real- or complex-valued matrix.
For a set $S$ of complex vectors of size $p$, define $\Realpart S 
\df\CBRbig{
  \bs\in\mathamsbb{R}^p
  \mid \bs+ \J \bt\in S \text{ for some $\bt\in\mathamsbb{R}^p$}
}$ and $S\cap\mathamsbb{R}^p\df
\CBRbig{\bs\in\mathamsbb{R}^p \mid {\bs+\J\bm{0}\in S}}$, where $\J = 
\sqrt{-1}$.
For $A \in \mathamsbb{C}^{M \times N}$, \label{eq:idenitites}
\begin{IEEEeqnarray}{c'c}
  \label{eq:nullspaceprop}
  \mathcal{N}(A^H) \cap \mathamsbb{R}^M =
  \mathcal{N}\PARENSs{ \underline{A}^T   }, &
   \Realpart\PARENSbig{\mathcal{R}(A)} =  \mathcal{R}( \underline{A} )
 \end{IEEEeqnarray}
are the \emph{real} null space and range of $\underline{A}^T$ and 
$\underline{A}$, respectively, where
\begin{IEEEeqnarray}{c}
  \label{eq:underlineA}
  \underline{A} \df \SBRbig{\Realpart A  \; \imag A } \in \mathamsbb{R}^{M 
  \times 2 N}.
\end{IEEEeqnarray}
If $\underline{A}$ in \eqref{eq:underlineA}
 has full row rank, we can define
\begin{IEEEeqnarray}{c}
  A^\ddagger \df A^H \SBRs{\Realpart(A A^H)}^{-1}
  \label{eq:Xi}
\end{IEEEeqnarray}
which reduces to $A^+$ for real-valued $A$.
The following are equivalent:
$\Realpart(\mathcal{R}\PARENSs{\Psi})=\mathamsbb{R}^p$,
$\mathcal{N}(\Psi^H) \cap \mathamsbb{R}^p=\{ \bm{0} \}$,
and
$d=p$, where
 \begin{IEEEeqnarray}{c}
    \label{eq:PsiRr}
    d \df   \dim\PARENSs{\Realpart\PARENSs{\mathcal{R}(\Psi)}} \leq 
    \min(p,2p').
  \end{IEEEeqnarray}
We can decompose $\Psi$ as
  \begin{IEEEeqnarray}{c}
    \Psi = F Z
    \label{eq:PsiFGgen}
  \end{IEEEeqnarray}
where $F \in \mathamsbb{R}^{p \times d}$ and  $Z \in \mathamsbb{C}^{d 
\times p'}$ with $\rank F = d$ and $\rank \underline{Z} = d$; 
$\underline{Z} = \SBRbig{\Realpart Z \; \imag Z} \in \mathamsbb{R}^{d 
\times 2 p'}$, consistent with the notation in \eqref{eq:underlineA}.  
Here, $\mathcal{R}\PARENSs{F}$ denotes the real range of the real-valued 
matrix $F$.
Clearly, $d\geq 1$ is of interest; otherwise $\Psi=0$.
Observe that  (see \eqref{eq:Xi})
\begin{subequations}
 \begin{IEEEeqnarray}{rCl}
  \label{eq:RePsiXi}
  \Realpart(\Psi Z^\ddagger)&=&F\\
  \label{eq:rangeF}
  \mathcal{R}\PARENSs{F}&=&\Realpart(\mathcal{R}\PARENSs{\Psi}).
\end{IEEEeqnarray}
\end{subequations}

The subdifferential of the indicator function 
$N_C(\bx)=\partial\mathbb{I}_C(\bx)$ is the \emph{normal cone to $C$ at 
$\bx$}
%\cite[Ch.~23, p.~215]{Rockafellar1970}, 
%\cite[Sec.~3.1]{BertsekasConvOptAlg2015}
\cite[Sec.~5.4]{BertsekasConvOptTheory2009}
and, by the definition of a cone, satisfies
\begin{IEEEeqnarray}{c"s}
  \label{eq:conescalingprop}
  N_C(\bx)=aN_C(\bx), & for any $a>0$.
\end{IEEEeqnarray}
Define
\begin{IEEEeqnarray}{c}
  \label{eq:Gs}
  G(s)\df\ccases{
    \CBRs{s/{|s|}}, & s\neq0\\
    \CBRs{w\in\mathamsbb{C} \mid {|w|\leq1}}, & s=0
  }
\end{IEEEeqnarray}
and its elementwise extension $G(\bs)$ for vector arguments $\bs$, which 
can be interpreted as twice the Wirtinger subdifferential of $\|\bs\|_1$ 
with respect to $\bs$ \cite{BouboulisSlavakisTheodoridis2012}.  Note that 
$\bs^HG(\bs)=\CBRs{\norm{\bs}_1}$, and, when $\bs$ is a real vector, 
$\Realpart\PARENSs{G(\bs)}$ is the subdifferential of $\norm{\bs}_1$ with 
respect to $\bs$ \cite[Sec.~11.3.4]{KabanavaRauhut2015}.

%Should we say that the model \eqref{eq:penalizedNLL}
%can handle fused lasso \cite{Tibshiranifusedlasso2005} as well, but fused 
%lasso has two regularization tuning parameters, so we would need to fix 
%one?
%I can see potential for more here, say find bounds for the two parameters 
%of fused lasso? Maybe the new students can try this?

%  For $\Psi\in\mathamsbb{C}^{p\times p'}$ and $\bx\in\mathamsbb{R}^p$, the 
%  subdifferential of $\norm{\Psi^H\bx}_1$ with respect to $\bx$ is  
%  \begin{IEEEeqnarray}{c}
%    \label{eq:subgradL1}
%    \partial_\bx \norm{\Psi^H \bx}_1 = \Realpart\PARENSbig{\Psi 
%    G\PARENSs{\Psi^H\bx}}
%  \end{IEEEeqnarray}
% which follows from
%  \begin{IEEEeqnarray}{rCl}
%    \label{eq:subgradAbs}
%    \partial_\bx|\bpsi_j^H\bx|
%    =
%    \Realpart\PARENSbig{\bpsi_j G(\bpsi_j^H\bx)}
%  \end{IEEEeqnarray}
%  where $\bpsi_j$ is the $j$th column of $\Psi$.  We obtain 
%  \eqref{eq:subgradAbs} by replacing the linear transform matrix in 
%  \cite[Prop.~2.1]{Wang2008TVcont} with $\SBRbig{\Realpart{\bpsi_j} \,\, 
%  \imag{\bpsi_j}}^T$.

\begin{lem}
  \label{lem:subd}
  For $\Psi\in\mathamsbb{C}^{p\times p'}$ and $\bx\in\mathamsbb{R}^p$, the 
  subdifferential of $\norm{\Psi^H\bx}_1$ with respect to $\bx$ is  
  \begin{IEEEeqnarray}{c}
    \label{eq:subgradL1}
    \partial_\bx \norm{\Psi^H \bx}_1 = \Realpart\PARENSbig{\Psi 
    G\PARENSs{\Psi^H\bx}}.
  \end{IEEEeqnarray}
\end{lem}
\begin{IEEEproof}
  \eqref{eq:subgradL1} follows from
  \begin{IEEEeqnarray}{rCl}
    \label{eq:subgradAbs}
    \partial_\bx|\bpsi_j^H\bx|
    =
    \Realpart\PARENSbig{\bpsi_j G(\bpsi_j^H\bx)}
  \end{IEEEeqnarray}
  where $\bpsi_j$ is the $j$th column of $\Psi$.  We obtain 
  \eqref{eq:subgradAbs} by replacing the linear transform matrix in 
  \cite[Prop.~2.1]{Wang2008TVcont} with $\SBRbig{\Realpart{\bpsi_j} \,\, 
  \imag{\bpsi_j}}^T$.
\end{IEEEproof}

We now use Lemma~\ref{lem:subd}
to formulate the necessary and sufficient conditions for 
$\bx\in\cX_u$:
\begin{subequations}
\begin{IEEEeqnarray}{rCl}
  \label{eq:optCond}
  \bm{0} &\in&
  u \Realpart\PARENSbig{\Psi G\PARENSs{\Psi^H \bx }}
  + \nabla\cL( \bx ) + N_C( \bx )
\end{IEEEeqnarray}
and $\bx\in Q$:
\begin{IEEEeqnarray}{rCl}
  \label{eq:bxStarCond}
  \bm{0}&\in& \Realpart\PARENSbig{\Psi G\PARENSs{\Psi^H\bx}}
  +N_C\PARENSs{\bx}
       %\\
    %\label{eq:due2Cone}
    %&=&u\Psi \SBRbigg{
    %  \frac{\Psi^+\nabla\cL\PARENS{\bx_u}}{u}
    %  +G\PARENSbig{\Psi^H\bx_u} + \Psi^+N_C\PARENS{\bx_u}
    %}
 %   \IEEEeqnarraynumspace
\end{IEEEeqnarray}
\end{subequations}
respectively.

%The signal vectors $\bx^\diamond\in\cX^\diamond$ satisfy both 
%\eqref{eq:optCond} and \eqref{eq:bxStarCond}.
%\\A: what if $U=\infty$?

%\cite{Chambolle200TV}

When the signal vector $\bx = \vect X$ corresponds to an image $X \in 
\mathamsbb{R}^{J\times K}$, its isotropic and anisotropic \gls{TV} 
regularizations correspond to
\cite[Sec.~2.1]{ChambollePock2016}
\begin{subequations}
  \label{eq:PsiTV}
  \begin{IEEEeqnarray}{rCl"s"}
    \label{eq:Psiiso}
    \Psi&=&\Psi_\text{v}+\J \Psi_\text{h}\in\mathamsbb{C}^{JK\times JK} & 
    (isotropic)
    \\
    \Psi&=&\bigl[\Psi_\text{v}\,\,\Psi_\text{h}\bigr]\in\mathamsbb{R}^{JK\times 
    2JK}
    & (anisotropic)
    \label{eq:Psiani}
  \end{IEEEeqnarray}
\end{subequations}
respectively, where
  $\Psi_\text{v}=I_K\otimes D^T(J)$ and
  $\Psi_\text{h}=D^T(K)\otimes I_J$
are the vertical and horizontal difference matrices (similar to those in  
\cite[Sec.~15.3.3]{BoydUnpublishedBook}),  and
\begin{IEEEeqnarray}{c}
  \label{eq:DL}
  D(L)\df\begin{bmatrix*}[r]
    1&   -1&      &      &  \\
     &    1&    -1&      &  \\
     &     &\ddots&\ddots&  \\
     &     &      &    1 &-1\\
    0&    0&\cdots& 0 & 0
   \end{bmatrix*} \in\mathamsbb{R}^{L\times L}
\end{IEEEeqnarray}
obtained by appending an all-zero row from below to
the $(L-1) \times L$ upper-trapezoidal matrix with first row 
$\SBRbig{1,-1,0,\dotsc,0}$;  note that $D(1)=0$.
Here, $d = JK-1$ and
\begin{IEEEeqnarray}{c}
  \label{eq:NPsiT}
  \mathcal{N}(\Psi^H)=\mathcal{R}(\bm{1})
\end{IEEEeqnarray}
for both the isotropic and anisotropic \gls{TV} regularizations.

The scenario where
\begin{IEEEeqnarray}{rCl}
  \label{eq:cond}
  \mathcal{N}(\Psi^H) \cap C \neq \emptyset
\end{IEEEeqnarray}
holds is of practical interest: then $Q = \mathcal{N}(\Psi^H) \cap C$ and 
$\bx^\diamond \in \cX^\diamond$ globally minimize the regularization term: 
$r\PARENSs{\bx^\diamond}=0$.  If \eqref{eq:cond} holds and $\bx^\diamond 
\in \cX^\diamond$, then $G(\Psi^H\bx^\diamond)=H$, where
\begin{IEEEeqnarray}{c}
  \label{eq:H}
  H\df\CBRbig{ \bw\in\mathamsbb{C}^{p' \times 1} \mid 
  \norm{\bw}_\infty\leq1}.
\end{IEEEeqnarray}
If, in addition to \eqref{eq:cond},
\begin{itemize}
  \item $d=p$,
    then  $\cX^\diamond=Q=\{\bm{0}\}$;
  \item $\mathcal{N}(\Psi^H) \cap \mathamsbb{R}^p=\mathcal{R}(\bm{1})$, 
    then   $Q= \mathcal{R}(\bm{1}) \cap C$ and
    $\bx^\diamond \in \cX^\diamond$ are constant signals of the 
    form $\bx^\diamond = \bm{1} x_0^\diamond, \, x_0^\diamond \in 
    \mathamsbb{R}$.
   % ???
   % $N_Q(\bx)=\mathcal{R}^\perp(\bm{1})+N_C(\bx)$ if $Q$ has more than one 
   % element.\ad{so you cannot make a general expression for all cases?}
\end{itemize}

In Section~\ref{sec:Udefprop}, we define and explain an upper bound $U$ on 
useful regularization constants $u$ and establish conditions under which signal 
sparsity regularization is \emph{irrelevant} and finite $U$ \emph{does not 
exist}.
We then present an optimization problem for finding $U$ when 
\eqref{eq:cond} holds (Section~\ref{sec:theorem}), develop a general 
numerical method for computing bounds $U$ (Section~\ref{sec:alg}), present 
numerical examples (Section~\ref{sec:NumEx}), and make concluding remarks 
(Section~\ref{sec:conclusion}).

\section{Upper Bound Definition and Properties}
\label{sec:Udefprop}

Define
\begin{IEEEeqnarray}{c}
  \label{eq:Udef}
  U \df \inf\CBRbig{{u \geq 0} \givens{ \cX_u \cap Q \neq \emptyset } }.
\end{IEEEeqnarray}
  If $\cX_u \cap Q = \emptyset$ for all $u$, then finite $U$ does not 
  exist, which we denote by $U=+\infty$.

We now show that, if $u \geq U$, then the
the set of minimum points $\cX_u$ of the objective function does not 
change.
%  The following lemma quantifies non-empty intersections $\cX_u\cap Q$.
\begin{rem}
  \label{rem:XuCapQ}
\begin{enumerate}[label=(\alph*)]
  \item
    {For any $u$, $\cX_u\cap Q=\cX^\diamond$ if and only if $\cX_u\cap 
    Q\neq\emptyset$.} \label{rem:XuCapQ1}
  \item  \label{rem:XuCapQ2}
    Assuming $\cX_U\cap Q \neq \emptyset$ for some $U \geq 0$, $\cX_u = 
    \cX^\diamond$ for $u > U$.
\end{enumerate}
\end{rem}

\begin{IEEEproof}
  We first prove~\ref{rem:XuCapQ1}.
  Necessity follows by the existence of $\cX^\diamond$; see  
  \eqref{eq:xdiamond}.
  We argue sufficiency by contradiction.  Consider any 
  $\bx_u\in\cX_u\cap Q$; i.e., $\bx_u$ minimizes both $f_u(\bx)$ and 
  $r(\bx)$.  If $\bx_u \notin \cX^\diamond$, there exists a $\by\in 
  \cX^\diamond$ with $\cL(\by)<\cL(\bx_u)$ that, by the definition of 
  $\cX^\diamond$, also minimizes $r(\bx)$.  Therefore, 
  $f_u(\by)=\cL(\by)+ur(\by)<f_u(\bx_u)$, which contradicts the assumption 
  $\bx_u\in\cX_u$.  Therefore, $\cX_u\cap Q\subseteq\cX^\diamond$.  If 
  there exists a $\bz\in \cX^\diamond\subseteq Q$ such that 
  $\bz\notin\cX_u$, then $f_u(\bz)>f_u(\bx_u)$ which, since both $\bz$ and 
  $\bx_u$ are in $Q$, implies that $\cL(\bz)>\cL(\bx_u)$ and contradicts 
  the definition of $\cX^\diamond$.  Therefore, 
  $\cX^\diamond\subseteq\cX_u$.

  We now prove~\ref{rem:XuCapQ2}. By~\ref{rem:XuCapQ1}, $\cX_U\cap 
  Q=\cX^\diamond$, which confirms~\ref{rem:XuCapQ2} for $u=U$.
  Consider
  now $u>U$, a $\by \in \cX_U\cap Q = \cX^\diamond$, and any $\bx 
  \in\cX_u$. Then,
  \begin{subequations}
    \label{eq:xyareOptimal}
    \begin{IEEEeqnarray}{rCl}
      \cL(\bx)+Ur(\bx)&\geq&\cL(\by)+Ur(\by)\\
      \cL(\by)+ur(\by)&\geq&\cL(\bx)+ur(\bx).
    \end{IEEEeqnarray}
  \end{subequations}
  By summing the two inequalities in \eqref{eq:xyareOptimal} and 
  rearranging, we obtain $r(\by) \geq r(\bx)$.  Since $\by \in Q$, $\bx$ is 
  also in $Q$; i.e., $\cX_u\subseteq Q$,
  which implies  $\cX_u=\cX^\diamond$ by \ref{rem:XuCapQ1}.
\end{IEEEproof}

As $u$ increases, $\cX_u$ moves gradually towards $Q$ and, according to the 
definition \eqref{eq:Udef}, $\cX_u$ and $Q$ do not intersect when $u<U$.  
Once $u=U$, the intersection of the two sets is $\cX^\diamond$, and, by
Remark~\ref{rem:XuCapQ}\ref{rem:XuCapQ2},
$\cX_u = \cX^\diamond$ for all $u > U$.

\subsection{Irrelevant Signal Sparsity Regularization}
%\label{sec:U0}

\begin{rem}
  \label{rem:noNeed4r}
  The following claims are equivalent:
  \begin{enumerate}[label=(\alph*)]
    \item \label{item:hasOne}
      $\cX^\diamond\cap\cX_0\neq\emptyset$; i.e., there exists an 
      $\bx^\diamond\in\cX^\diamond$ such that
      \begin{IEEEeqnarray}{rCl}
        \label{eq:noNeed4r}
        \bm{0}\in\nabla\cL\PARENSs{\bx^\diamond}+N_C\PARENSs{\bx^\diamond};
      \end{IEEEeqnarray}
    \item \label{item:hasAll}
      %\eqref{eq:noNeed4r} holds for all $\bx^\diamond \in \cX^\diamond$; 
      $\cX^\diamond\subseteq\cX_0$;
      and
    \item \label{eq:Uzero}
      $U=0$; i.e., $\cX_0\cap Q\neq\emptyset$.
  \end{enumerate}
%    Thus, when \eqref{eq:noNeed4r} holds, signal sparsity regularization 
%    with $u>0$ does not affect $\cX_u$, which is equal to 
%    $\cX^\diamond=\cX_0\cap Q$.
\end{rem}
\begin{IEEEproof}
  \ref{eq:Uzero} follows from~\ref{item:hasOne} because 
  $\cX^\diamond\subseteq Q$.  \ref{item:hasAll} follows from~\ref{eq:Uzero} 
  by applying Remark~\ref{rem:XuCapQ}\ref{rem:XuCapQ1} to obtain $\cX_0\cap 
  Q=\cX^\diamond$, which implies \ref{item:hasAll}.  
  Finally,~\ref{item:hasAll} implies~\ref{item:hasOne}.
\end{IEEEproof}

%What will happen with \eqref{eq:noNeed4r}
% if $C=\mathamsbb{R}^{p \times 1}$?
%
%What will happen with \eqref{eq:noNeed4r} if $\bx^\diamond \in \interior 
%C$?
%\\A: then $\nabla\cL(\bx^\diamond)=\bm{0}$.  This includes the case above.

Having $\nabla\cL(\bx^\diamond)=\bm{0}$ for at least one 
$\bx^\diamond\in\cX^\diamond$ implies \eqref{eq:noNeed4r} and is therefore 
a stronger condition than \eqref{eq:noNeed4r}.

\begin{exa}
  \label{ex:zeroU}
  Consider $\cL(\bx)=\norm{\bx}_2^2$ and $C=\CBRbig{
    \bx\in\mathamsbb{R}^2 \mid \norm{\bx-\bm{1}_{2\times 1}}_2\leq1
  }$.
  (Here, $\cL(\bx)$ could correspond to the Gaussian measurement model with 
  measurements equal to zero.)
  Since $C$ is a circle within $\mathamsbb{R}_+^2$, the objective functions 
  for the identity ($\Psi=I_2$) and 1D \gls{TV} 
  %($\Psi=\SBRbig{[1, -1]^T,\bm{0}_{2\times1}}$, corresponding to $J=2$ and 
  %$K=1$ in \eqref{eq:Psiiso})
  sparsifying transforms are
  \begin{subequations}
    \begin{IEEEeqnarray}{rCl's}
      f_u(\bx)&=&x_1^2+x_2^2 + {u}(x_1+x_2)+\mathbb{I}_C(\bx), & (identity)  
      \\
      f_u(\bx)&=&x_1^2+x_2^2 + {u}|x_1-x_2|+\mathbb{I}_C(\bx), & (1D TV)
  \IEEEeqnarraynumspace
    \end{IEEEeqnarray}
  \end{subequations}
  respectively, where $\mathcal{X}_u = \mathcal{X}^\diamond=Q=\{ 
  \bx^\diamond \}$
  and
  $\bx^\diamond = \PARENSbig{1-{\sqrt{2}}/{2}}\bm{1}$.
  Here, 
  $\nabla\cL\PARENSs{\bx^\diamond}=\PARENSs{2-\sqrt{2}}\bm{1}_{2\times 1}$ 
  and $N_C\PARENSs{\bx^\diamond}=\CBRs{a\bm{1} \mid a\leq0}$, which 
  confirms that \eqref{eq:noNeed4r} holds.
%$N_C\PARENSbig{\bx^\diamond}=\CBR{ \bx \in \mathamsbb{R}^2 \mid   \bx = 
%a\bm{1}, a\leq0}$.
\end{exa}

\subsection{Condition for Infinite $U$ and Guarantees for Finite $U$}
%\label{sec:Uinfty}

\begin{rem}
  \label{rem:Uinfty}
  If there exists $\bx^\diamond\in\cX^\diamond$ such that
  \begin{IEEEeqnarray}{rCl}
    \label{eq:Uinftycond}
    \SBRs{\nabla\cL\PARENSs{\bx^\diamond}+N_C\PARENSs{\bx^\diamond}}
    \cap
    \Realpart(\mathcal{R}\PARENSs{ \Psi })=\emptyset.
  \end{IEEEeqnarray}
  then $U=+\infty$.  When \eqref{eq:cond}
  holds, the reverse is also true with a stronger claim: $U=+\infty$ 
  implies \eqref{eq:Uinftycond} for all $\bx^\diamond\in\cX^\diamond$.
\end{rem}
\begin{IEEEproof}
  First, we prove sufficiency by contradiction.  If a finite $U$ exists, 
  then $\cX^\diamond\subseteq\cX_u$ for all $u\geq U$.  Therefore, 
  \eqref{eq:optCond} holds with $\bx$ being any 
  $\bx^\diamond\in\cX^\diamond$, which contradicts \eqref{eq:Uinftycond}.

  In the case where \eqref{eq:cond} holds, we prove the necessity by 
  contradiction.
  If \eqref{eq:Uinftycond} does not hold for all 
  $\bx^\diamond\in\cX^\diamond$, there exist $\bt\in N_C(\bx^\diamond)$ and 
  $\bw\in\mathamsbb{C}^{p'}$ such that
  \begin{IEEEeqnarray}{c}
    \label{eq:containsZero}
    \bm{0} = \nabla\cL(\bx^\diamond)+\Realpart\PARENSs{\Psi\bw}+\bt.
  \end{IEEEeqnarray}
  Since \eqref{eq:cond} holds, $\Psi^H \bx^\diamond = \bm{0}$ and 
  $G(\Psi^H\bx^\diamond)=H$; see \eqref{eq:H}.  When 
  $u\geq\norm{\bw}_\infty$, $\bw\in uH$ and $\Realpart\PARENSs{\Psi\bw}\in
  u\Realpart\PARENSbig{\Psi G\PARENSs{\Psi^H\bx^\diamond}}$.   Therefore, 
  \eqref{eq:optCond} holds at $\bx=\bx^\diamond$ for all 
  $u\geq\norm{\bw}_\infty$, which contradicts $U=+\infty$.
% Since $U=+\infty$, $\cX_u\cap Q=\emptyset$ for all finite $u$.  Because 
% $\cX^\diamond\subseteq Q$, $\cX_u\cap\cX^\diamond=\emptyset$, which means 
% \eqref{eq:Uinftycond} holds for all $\bx^\diamond\in\cX^\diamond$.
\end{IEEEproof}

%\eqref{eq:Uinftycond2} is not possible. Since $\cX^\diamond$ is already 
%defined, left hand side of \eqref{eq:Uinftycond2} is well defined and the 
%problem is not here.  The problem is to let \eqref{eq:optCond} hold.  For 
%\eqref{eq:Uinftycond} not hold, means that you can always find some $\bw$ 
%such that
%\begin{equation}
%  \bm{0}\in\SBRs{\nabla\cL\PARENSs{\bx^\diamond}+N_C\PARENSs{\bx^\diamond}}
%  +\Realpart( \Psi \bw)
%\end{equation}
%But that does not mean \eqref{eq:optCond} because $G(\Psi^H\bx^\diamond)$ 
%does not necessary contain such a $\bw$.

%The following example illustrates an interesting case where $U=+\infty$.
\begin{exa}
  \label{ex:Uninfinite}
  Consider $\cL(\bx)=x_1+\mathbb{I}_{\mathamsbb{R}_+}(x_1)$, $\Psi=I_2$, 
  and
  $C=\CBRbig{
    \bx\in\mathamsbb{R}^2 \mid \norm{\bx-\bm{1}_{2\times 1}}_2\leq1
  }$.
  (Here, $\cL(\bx)$ could correspond to the
  $\Poisson(x_1)$ measurement model with measurement equal to zero.)
  Since $C$ is a circle within $\mathamsbb{R}_+^2$, the objective function 
  is
  \begin{IEEEeqnarray}{rCl}
    f_u(\bx)=(1+u)x_1+ux_2+\mathbb{I}_C(\bx)
  \end{IEEEeqnarray}
  with $\mathcal{X}_u = \{ \bx_u \}$,  $\mathcal{X}^\diamond = Q = \{ 
  \bx^\diamond \}$,
and
\begin{subequations}
  \begin{IEEEeqnarray}{rCl}
    \label{eq:xustarex1}
    \bx_u &=& \bm{1}_{2 \times 1}-\frac{1}{\sqrt{2+{2}/{u}+{1}/{u^2}}}
    \begin{bmatrix}
      1+{1}/{u}\\
      1
    \end{bmatrix}\\
    \label{eq:xstarex1}
    \bx^\diamond &=& \PARENSbig{1-{\sqrt{2}}/{2}}\bm{1}_{2\times 1}
  \end{IEEEeqnarray}
\end{subequations}
which implies $U=+\infty$, consistent with
the observation that $\cX_u \cap Q = \emptyset$.  Here, \eqref{eq:cond} is 
not satisfied:  \eqref{eq:Uinftycond} is only a sufficient condition for 
$U=+\infty$ and does not hold in this example.   
%due to $\Psi^H\bx^\diamond\neq\bm{0}$.
%Note also that $Q= \cX^\diamond$ in this example.

\end{exa}

\begin{exa}
  \label{ex:UninfiniteTV}
  Consider $\cL(\bx)=\norm{\bx}^2_2$, 1D \gls{TV} sparsifying transform 
  with $\Psi=D^T(2)$, and
  $C=\CBRbig{
    \bx\in\mathamsbb{R}^2 \mid \normbig{\bx-\SBRbig{2, \; 0}^T}_2^2\leq2
  }$.
  Since $C$ is a circle with $x_1-x_2\geq0$, the objective function is
  \begin{subequations}
    \begin{IEEEeqnarray}{rCl}
      f_u(\bx)&=&\norm{\bx}_2^2+u|x_1-x_2|+\mathbb{I}_C(\bx)
      \\
      &=&
      \norm{\bx-\tfrac{1}{2}[u\,\,{-u}]^T}_2^2
      -{u^2}/{2}+\mathbb{I}_C(\bx)
    \end{IEEEeqnarray}
  \end{subequations}
  with $\mathcal{X}_u = \CBRbig{
    \SBRbig{
      2-(1+{4}/{u})/q(u), \;
      1/q(u)
  }^T }$, $q(u) \df \sqrt{1+{4}/{u}+{8}/{u^2}}$,
  and $\mathcal{X}^\diamond = Q = \{ \bm{1}_{2\times 1} \}$, which implies 
  $U=+\infty$.  Since \eqref{eq:cond} holds in this example, 
  \eqref{eq:Uinftycond} is necessary and sufficient for $U=+\infty$.  Since  
  $-\bm{1}^T\nabla\cL\PARENSs{\bx^\diamond}=-4$ and 
  $N_C\PARENSs{\bx^\diamond}=\CBRs{(-a,a)^T \mid a\geq0}$,  
  \eqref{eq:Uinftycond} holds.
%  In this example,
%$Q= \cX^\diamond =\CBRs{\bx^\diamond}$.
\end{exa}

\subsubsection{Two cases of finite $U$}
%\label{sec:twocasesfiniteU}

If  $d=p$ and \eqref{eq:cond} holds, then $U$ must be finite:  in this 
case,  condition \eqref{eq:Uinftycond} in Remark~\ref{rem:Uinfty} cannot 
hold, which is easy to confirm by substituting 
$\Realpart(\mathcal{R}\PARENSs{\Psi})=\mathamsbb{R}^p$ into 
\eqref{eq:Uinftycond}.

%???A: I think we should use the concept ``relative interior'' with the 
%affine hyperplane set to $Q+\PARENSbig{\mathcal{N}(\Psi^H) \cap 
%\mathamsbb{R}^p}$.
%
%I got somewhere else.  But also find it in the reference that you mentioned below.  http://www2.isye.gatech.edu/~nemirovs/OPTIII_LectureNotes.pdf
%
%??? ok, seems like Boyd and Bersekas and Rockafellar define is, so you can 
%cite one of them.
%
%This is section title in Boyd:
%2.1.3 Affine dimension and relative interior
%
%Bertsekas D.P.-Convex_optimization_algorithms---2015
%is full of relative interior statements
%
%This is section title in Bertsekas convex_optimization_theory 2009
%1.3 RELATIVE INTERIOR AND CLOSURE
%which we cite already

%??? A: I was thinking to make the condition weaker.  But it seems that it 
%has to be the general interior.  $\cX^\diamond\cap \relint C$ has to have 
%nonempty to make \eqref{eq:cond} holds by choosing the affine space to be 
%$\Realpart\mathcal{R}(\Psi)$,  while we also want $\cX^\diamond\cap 
%\relint C\neq\emptyset$ with affine space to be $\mathcal{N}(\Psi^H)$ to 
%make $N_C(\bx^\diamond)=\{0\}$.  Therefore, we need to have it be the 
%general interior.

$U$ must also be finite if
\begin{IEEEeqnarray}{c}
    \label{eq:condint}
  %\label{eq:XdiamondcapintCnotempty}
  \cX^\diamond \cap \interior C\neq\emptyset.
\end{IEEEeqnarray}
Indeed, \eqref{eq:condint}
implies \eqref{eq:cond} and that for $\bx^\diamond \in \cX^\diamond\cap 
\interior C$,
\begin{subequations}
  \begin{IEEEeqnarray}{rCl}
    \label{eq:NCzero}
    N_C\PARENSs{\bx^\diamond} &=& \{\bm{0}\}\\
     \label{eq:dueToInInteriorC}
  \nabla\cL(\bx^\diamond) &\in& \Realpart\PARENSs{\mathcal{R}(\Psi)} 
\end{IEEEeqnarray}
\end{subequations}
and hence \eqref{eq:Uinftycond} cannot hold upon substituting 
\eqref{eq:NCzero} and \eqref{eq:dueToInInteriorC}.  Here, 
\eqref{eq:dueToInInteriorC} follows from 
$\bm{0}\in\nabla\cL(\bx^\diamond)+N_Q(\bx^\diamond)$, the condition for 
optimality of the optimization problem $\min_{\bx\in Q}\cL(\bx)$ that 
defines $\cX^\diamond$, by using the fact that $N_Q(\bx^\diamond) = 
\Realpart\PARENSs{\mathcal{R}(\Psi)}$ when $\bx^\diamond \in 
\cX^\diamond\cap \interior C$.

%A: this is difficult.   I can only come up one extreme example which is
%$\bm{0}\notin C$, $N_C(\bx^\diamond)$ has only one direction, 
%$\Psi^H\bx^\diamond$ has no zero elements, and $\nabla\cL(\bx^\diamond)$ is 
%not collinear with $N_C(\bx^\diamond)$ and $\Psi^H\bx^\diamond$.
%{??? well we can at least comment what happens with the above remark in 
%this case and also anything that you have, such as this example. maybe you 
%can say what is difficult}
%\\A: actually above is exact Example 2.  It is hard because we may make 
%some zeros in $\Psi^H\bx^\diamond$ and $N_C(\bx^\diamond)$ has more 
%directions, but the condition on $\nabla\cL(\bx^\diamond)$ will be very 
%difficult to describe.

%???
%The point here is that since we assume $\bx^\diamond$ exists, we have  
%$-\nabla\cL(\bx^\diamond)\in N_Q(\bx^\diamond)$, where $Q$ is defined in 
%\eqref{eq:Q}.  Due to the definition of $Q$, we have that 
%$\Realpart\PARENSbig{\Psi 
%  G\PARENSs{\Psi^H\bx^\diamond}}+N_C(\bx^\diamond)\subseteq 
%  N_Q(\bx^\diamond)$.
%Then, the point is when does the above two sets are equal to each other.
%I have worked on it for a few days without finding the condition.

If \eqref{eq:condint} holds then,  by Remark~\ref{rem:noNeed4r}, $U=0$ if 
and only if $\nabla\cL\PARENSs{\bx^\diamond}=\bm{0}$.

\section{Bounds When \eqref{eq:cond} Holds}
\label{sec:theorem}

We now present an optimization problem for finding $U$ when \eqref{eq:cond} 
holds.
%and the regularization term can be globally minimized by a feasible $\bx$.

\begin{thm}
  \label{thm:U}
  Assume that \eqref{eq:cond} holds and that the convex \gls{NLL} 
  $\cL(\bx)$ is differentiable within $\cX^\diamond$.  Consider the 
  following optimization problem:
  \begin{subequations}
    \label{eq:Uth}
    \begin{IEEEeqnarray}{s'c+l}
      \label{eq:Uthminproblem}
      (P$_0$): & U_0(\bx^\diamond) = & \kern-15pt
      \min_{\ba \in \mathamsbb{R}^p, \, \bt \in \mathamsbb{C}^{p'}}
      \norm{\bp(\bx^\diamond,\ba,\bt)}_\infty
      \\
      & &
      \text{subject to} \label{eq:ainNC}
      \kern10pt  \ba \in N_C\PARENSs{\bx^\diamond}\\
      \label{eq:gradrealpartRPsi}
      & &
      \kern48pt \nabla\cL\PARENSs{\bx^\diamond}+\ba
      \in
      \mathcal{R}\PARENSs{F}
      \IEEEeqnarraynumspace
    \end{IEEEeqnarray}
  \end{subequations}
  with
  \begin{IEEEeqnarray}{c}
    \label{eq:bp}
    \bp(\bx,\ba,\bt)
    \df \bt + 
    Z^\ddagger\CBRbig{F^+\SBRs{\nabla\cL\PARENSs{\bx}+\ba}-\Realpart(Z\bt)}.
  \end{IEEEeqnarray}
  Then,
  $U_0(\bx^\diamond)=U$ for all $\bx^\diamond\in\cX^\diamond$ and $U$  in 
  \eqref{eq:Udef}.
\end{thm}
Here, 
%
%$U=0$ if and only if \eqref{eq:noNeed4r} in Remark~\ref{rem:noNeed4r} 
%holds and 
%
$U=+\infty$ if and only if the constraints in  \eqref{eq:ainNC} and 
\eqref{eq:gradrealpartRPsi} cannot be satisfied for any $\ba$, which is 
equivalent to $\bx^\diamond\in\cX^\diamond$ satisfying 
\eqref{eq:Uinftycond} in Remark~\ref{rem:Uinfty}.

\begin{IEEEproof}
  \label{app:proofofU}
  Observe that $G(\Psi^H\bx^\diamond)=H$ for all  
  $\bx^\diamond\in\cX^\diamond$ and
  \begin{IEEEeqnarray}{c}
    \label{eq:RePsibp}
    \Realpart\PARENSbig{\Psi\bp(\bx,\ba,\bt)}
    =\nabla\cL\PARENSs{\bx}+\ba.
  \end{IEEEeqnarray}
  due to \eqref{eq:cond} and \eqref{eq:RePsiXi}, respectively.

  We first prove that $\cX^\diamond \subseteq \cX_u$ if $u\geq 
  U_0(\bx^\diamond)$.
%if $u\geq U_0(\bx^\diamond)$, then $\bx^\diamond\in\cX_u$
%  for all $\bx^\diamond\in\cX^\diamond$.
%  why are you using this instead of sets and intersections of sets?  sets 
%  is in line with the rest of the paper. you can use $\bx^\diamond$ only 
%  when needed
%??? Isn't this $\cX^\diamond \subseteq \cX_u$?
  Consider \emph{any} $\bx^\diamond\in\cX^\diamond$ and denote by 
  $\PARENSs{\tilde{\ba},\tilde{\bt}}$ a pair $\PARENSs{\ba,\bt}$ that 
  solves the minimization problem (P$_0$).  Since $u\geq 
  U_0(\bx^\diamond)$, there exists an $\tilde{\bh}\in H$ such that 
  $\bp\PARENSs{\bx^\diamond,\tilde{\ba},\tilde{\bt}}+u\tilde{\bh}=\bm{0}$.  
  Using \eqref{eq:RePsibp}, we obtain
  \begin{IEEEeqnarray}{c}
    \bm{0}=\Realpart\CBRs{
      \Psi\SBRs{\bp\PARENSs{\bx^\diamond,\tilde{\ba},\tilde{\bt}}
        +u\tilde{\bh}
      }
    } 
    =u\Realpart\PARENSs{\Psi\tilde{\bh}}+\nabla\cL(\bx^\diamond)+\tilde{\ba}
    \notag \\
  \end{IEEEeqnarray}
  which implies $\bx^\diamond\in\cX_u$ according to 
  \eqref{eq:optCond}.

  Second, we prove that if $u< U_0(\bx^\diamond)$ for any 
  $\bx^\diamond\in\cX^\diamond$, then $\cX^\diamond\cap\cX_u=\emptyset$. We 
  employ proof by contradiction.  Suppose 
  $\cX^\diamond\cap\cX_u\neq\emptyset$; then, there exists an 
  $\bx^\diamond\in\cX^\diamond\cap\cX_u$. According to \eqref{eq:optCond}, 
  there exist an $\check{\bh}\in H$ and an $\check{\ba}\in 
  N_C(\bx^\diamond)$ such that 
  $\bm{0}=u\Realpart\PARENSs{\Psi\check{\bh}}+\nabla\cL(\bx^\diamond)+\check{\ba}$.  
  Using \eqref{eq:RePsibp}, we have
  \begin{IEEEeqnarray}{rCl}
    \label{eq:Due2Xu}
    \bm{0}
    &=&
    \Realpart\PARENSbig{
      \Psi\SBRs{u\check{\bh}+\bp\PARENSs{\bx^\diamond,\check{\ba},-u\check{\bh}}}
    }.
  \end{IEEEeqnarray}
  Note that
  \begin{IEEEeqnarray}{rCl}
    u\check{\bh}+\bp\PARENSs{\bx^\diamond,\check{\ba},-u\check{\bh}}
    &=&
    Z^\ddagger\CBRbig{F^+\SBRs{\nabla\cL\PARENSs{\bx^\diamond}+\check{\ba}}+u\Realpart(Z\check{\bh})}.  
    \notag
    \\
    \label{eq:id}
  \end{IEEEeqnarray}
  Inserting \eqref{eq:id} into \eqref{eq:Due2Xu} and using 
  \eqref{eq:RePsiXi} and the fact that $F$ has full column rank leads to 
  $\bm{0}={F^+\SBRs{\nabla\cL\PARENSs{\bx^\diamond}+\check{\ba}}+u\Realpart(Z\check{\bh})}$; 
  thus
  \begin{IEEEeqnarray}{rCl}
    \bm{0}
    &=&
    u\check{\bh}+\bp\PARENSs{\bx^\diamond,\check{\ba},-u\check{\bh}}.
  \end{IEEEeqnarray}
  Now, 
  %
  %take $u\check{\bh}$ to the left hand side, take infinity norm on both 
  %sides, 
  %
  rearrange and use the fact that $\norm{\check{\bh}}_\infty\leq1$  (see 
  \eqref{eq:H}) to obtain
  \begin{IEEEeqnarray}{c}
    \label{eq:pid}
    \norm{\bp\PARENSs{\bx^\diamond,\check{\ba},-u\check{\bh}}}_\infty
    = u\norm{-\check{\bh}}_\infty
    \leq u
    < U_0(\bx^\diamond)
  \end{IEEEeqnarray}
  which contradicts \eqref{eq:Uth}, where $U_0(\bx^\diamond)$ is the 
  minimum.

  Finally, we prove by contradiction that $U_0(\bx^\diamond)$ is invariant 
  within $\cX^\diamond$ if $\cX^\diamond$ has more than one element.  
  Assume that there exist $\bx^\diamond_1, \bx^\diamond_2 \in \cX^\diamond$ 
  and $u$ such that $U_0(\bx^\diamond_1)\leq u<U_0(\bx^\diamond_2)$.  We 
  obtain contradictory results: $\bx^\diamond_1\in\cX_u$ and 
  $\cX^\diamond\cap\cX_u\neq\emptyset$ because $u\geq U_0(\bx^\diamond_1)$ 
  and $u<U_0(\bx^\diamond_2)$, respectively.  Therefore, 
  $U=U_0(\bx^\diamond)$ is invarant to $\bx^\diamond\in\cX^\diamond$.

 The constraints on $\ba$ in \eqref{eq:ainNC} and 
 \eqref{eq:gradrealpartRPsi} are equivalent to stating that 
 \eqref{eq:Uinftycond} does not hold for any $\bx^\diamond \in 
 \mathcal{X}^\diamond$; see also \eqref{eq:rangeF}.  If an $\ba$ does not 
 exist that satisfies these constraints, \eqref{eq:Uinftycond} holds and 
 $U=+\infty$ according to Remark~\ref{rem:Uinfty}.
\end{IEEEproof}

We make a few observations: (P$_0$) is a linear programming problem with 
linear constraints and can be solved using CVX \cite{cvx} and Matlab's 
optimization toolbox upon identifying $N_C(\bx^\diamond)$ and 
$\mathcal{R}(F)$ in \eqref{eq:ainNC} and \eqref{eq:gradrealpartRPsi}, 
respectively.  Theorem~\ref{thm:U} requires differentiability of the 
\gls{NLL} only at $\bx=\bx^\diamond  \in \cX^\diamond$.  If $\Psi$ is real, 
then $Z$ is real as well, the optimal $\bt$ in (P$_0$) has zero imaginary 
component and the corresponding simplified version of Theorem~\ref{thm:U} 
follows and requires optimization in (P$_0$) with respect to real-valued 
$\bt \in \mathamsbb{R}^{p'}$.

If $\Psi$ is real and $d = p'$, then we can select $Z=I$, which leads to 
$Z^\ddagger=I$ and cancellation of the variable $\bt$ in 
\eqref{eq:Uthminproblem} and simplification of (P$_0$).

We now specialize Theorem~\ref{thm:U} to two cases with finite $U$.

\begin{cor}[$d=p$]
  \label{cor:deqp}
  If $d=p$ and if \eqref{eq:cond} holds, then $U$ in \eqref{eq:Udef} can be 
  computed as
  \begin{IEEEeqnarray}{c}
    \label{eq:Uthdeqp}
    U = \min_{\ba \in  N_C\PARENSs{\bm{0}}, \, \bt \in \mathamsbb{C}^{p'}}
    \normbig{
      \bt
      +
      \Psi^\ddagger\SBRs{\nabla\cL\PARENSs{\bm{0}}+\ba-\Realpart(\Psi\bt)}  
    }_\infty.  \IEEEeqnarraynumspace
  \end{IEEEeqnarray}
\end{cor}
\begin{IEEEproof}
  Theorem~\ref{thm:U} applies, $\mathcal{X}^\diamond=\{\bm{0}\}$, and $U$ 
  must be finite.  Setting $F=I$ in \eqref{eq:Uth} leads to 
  \eqref{eq:Uthdeqp}.
\end{IEEEproof}
If $C=\mathamsbb{R}_+^p$, then $N_C(\bm{0})=\mathamsbb{R}_-^p$ and the 
condition $\ba \in  N_C\PARENSs{\bm{0}}$ reduces to $\ba \preceq \bm{0}$.

% \begin{IEEEproof}
%If $d=p$ and \eqref{eq:cond} hold,
% then $\bx^\diamond=\bm{0}$ 
%   
%   
%   $d=p$ implies that
%\begin{itemize}
%  \item $\mathcal{N}(\Psi^T) \cap \mathamsbb{R}^p = \{ \bm{0} \}$,
%which, together with \eqref{eq:cond} implies
%  $\mathcal{N}(\Psi^T) \cap C = \{ \bm{0} \}$
%and
%$\bx^\diamond=\bm{0}$;
%
%\item we can select $F=I$ and remove the constraint 
%  \eqref{eq:gradrealpartRPsi}. \hfil \IEEEQEDhere  \end{itemize}
% \end{IEEEproof}

\begin{cor}[$\cX^\diamond \cap \interior C\neq\emptyset$]
  \label{cor:xstarintC}
  If \eqref{eq:condint} holds, then $U$ in \eqref{eq:Udef} can be computed 
  as
  \begin{IEEEeqnarray}{rCl}
    \label{eq:Uthreqp}
    U&=&\min_{\bt \in \mathamsbb{C}^d}
    \normbig{ \bt + Z^\ddagger
      \SBRs{F^+ \nabla\cL\PARENSs{\bx^\diamond} - \Realpart(Z \bt) }  
    }_\infty
  \end{IEEEeqnarray}
\end{cor}
with any $\bx^\diamond \in \cX^\diamond \cap \interior C$.

\begin{IEEEproof}
  Thanks to \eqref{eq:condint}, \eqref{eq:cond} and 
  \eqref{eq:NCzero}--\eqref{eq:dueToInInteriorC} are satisfied, 
  Theorem~\ref{thm:U} applies,
  $U$ must be finite, and $\ba=\bm{0}$ (by \eqref{eq:NCzero}). By using 
  these facts,  we simplify \eqref{eq:Uth} to obtain \eqref{eq:Uthreqp}.
%
%
%   Since \eqref{eq:Uinftycond} does not hold, we have
%\begin{IEEEeqnarray}{rCl}
%    \label{eq:Uinftycond2}
%    \SBRs{\nabla\cL\PARENSs{\bx^\diamond}+N_C\PARENSs{\bx^\diamond}}
%    \cap \mathcal{R}\PARENSs{ F } \neq \emptyset.
%  \end{IEEEeqnarray}  Substituting $N_C\PARENSs{\bx^\diamond}=\{\bm{0}\}$
%   into \eqref{eq:Uinftycond2} implies that
%   \eqref{eq:gradrealpartRPsi} always holds.
\end{IEEEproof}

%??? What about $d=p$ and $\bm{0} \in \interior C$? This seems like a mix 
%of the above corollaries and perhaps a good example (and more general than 
%the one below).
%\\A: in this case remove $\ba$ from \eqref{eq:Uthdeqp}.

If $d=p$ and $\bm{0} \in \interior C$, then both Corollaries~\ref{cor:deqp}
and \ref{cor:xstarintC} apply and the upper bound $U$ can be obtained by 
setting $\ba=\bm{0}$  and $N_C(\bm{0})=\{\bm{0}\}$ in
\eqref{eq:Uthdeqp} or
by setting  $\bx^\diamond = \bm{0}$ and  $F=I$ in  \eqref{eq:Uthreqp}.

\begin{exa}
  \label{ex:nonnegsig}
  Consider a real invertible $\Psi \in \mathamsbb{R}^{p\times p}$.
  \begin{enumerate}[label=(\alph*)]
    \item If $C=\mathamsbb{R}_+^p$,
      Corollary~\ref{cor:deqp} applies and \eqref{eq:Uthdeqp} becomes
  \begin{subequations}
  \begin{IEEEeqnarray}{c}
    U=
    \min_{\ba \preceq \bm{0}}
    \norm{\Psi^{-1} \SBRs{\nabla\cL\PARENSs{\bm{0}} + \ba }}_\infty.
    \label{eq:UCnonnegZsquare}
  \end{IEEEeqnarray}
  In this case, $U=0$ and signal sparsity regularization is irrelevant if 
  $\nabla\cL(\bm{0}) \succeq \bm{0}$, which follows by inspection from  
  \eqref{eq:UCnonnegZsquare}, as well as from \eqref{eq:noNeed4r} in 
  Remark~\ref{rem:noNeed4r}. If $\Psi=I$, \eqref{eq:UCnonnegZsquare}
  further reduces to $U=-\min\PARENSbig{0,\min_i[\nabla\cL(\bm{0})]_i}$.
  \item
  \label{ex:Gausslinmodel}
  If $\bm{0} \in \interior C$, Corollaries~\ref{cor:deqp}
  and \ref{cor:xstarintC} apply and the bound $U$ simplifies to
  \begin{IEEEeqnarray}{c}
    \label{eq:UinvertiblerealPsizerointC}
    U = \norm{\Psi^{-1} \nabla\cL\PARENSs{\bm{0}}  }_\infty.
  \end{IEEEeqnarray}
\end{subequations}
%  For linear measurement model with white Gaussian noise and scaled 
%  \gls{NLL}:
%  \begin{subequations}
%\begin{IEEEeqnarray}{c"c}
%  \cL(\bx) = \tfrac{1}{2} \|\by-\Phi\bx\|_2^2,
%& \Phi \in \mathamsbb{R}^{N \times p}???
%      \label{eq:gaussianL}
%    \end{IEEEeqnarray} 
%    \eqref{eq:UinvertiblerealPsizerointC} becomes
%\begin{IEEEeqnarray}{c"c}
%      \label{eq:U0linGauss}
%      U = \norm{\Psi^{-1}\Phi^T\by}_\infty.
%    \end{IEEEeqnarray}
    For $\Psi=I$ and a linear measurement model with white Gaussian noise, 
    \eqref{eq:UinvertiblerealPsizerointC} reduces to the expressions
    in \cite[eq.~(4)]{KimBoyd2007} and \cite[Sec.~III]{Wright2009SpaRSA}, 
    used in \cite{Wright2009SpaRSA} to design its continuation scheme;  
    \cite{KimBoyd2007} and \cite{Wright2009SpaRSA} also assume 
    $C=\mathamsbb{R}^p$.
 % \end{subequations}

\end{enumerate}
\end{exa}

%isotropic TV:
%??? even it can be written in certain form, it still has $\bt$ and $\ba$ in 
%the expression.\ad{I think it has only $\ba$ in the expression. You can 
%comfortably assume $\bx^\diamond \in \interior C$, so the above corollary 
%should hold}
%
%%??? What can you say if $C=\mathamsbb{R}_+^p$ and  $\bx^\diamond \neq 
%%\bm{0}$, useful for TV? If this helps, you can assume $\bx^\diamond$ 
%%constant. in TV case, if  $\bx^\diamond \neq \bm{0}$ then it must be in 
%%interior of $C$.  So, $N_C\PARENSs{\bx^\diamond}=\{\bm{0}\}$.
%%
%??? If
%\begin{IEEEeqnarray}{c}
%    \label{eq:PsiR1}
%    \mathcal{N}(\Psi^T) \cap \mathamsbb{R}^p=\mathcal{R}(\bm{1}),
%  \end{IEEEeqnarray}
%then $d=p-1$  and
%$\mathcal{R}(F) = \mathcal{R}^\perp(\bm{1})$
%??? relevant for isotropic TV

\begin{exa}[One-dimensional TV regularization]
  \label{ex:1DTV}
  Consider 1D \gls{TV} regularization with $\Psi=D^T(p) \in 
  \mathamsbb{R}^{p \times p}$ obtained by setting $K=1,J=p$ in 
  \eqref{eq:Psiiso}; note that $d=p-1$.   Consider a constant signal 
  $\bx^\diamond = \bm{1} x_0^\diamond \in \cX^\diamond$. Then 
  Theorem~\ref{thm:U}
  applies and yields
  \begin{subequations}
  \begin{IEEEeqnarray}{rCl}
    \label{eq:UTV}
    U=  \min_{\ba\in N_C( \bm{1} x_0^\diamond )}  \max_{1 \leq j < p}
    \absbigg{\sum_{i=1}^j \SBRbig{   \nabla\cL\PARENSs{\bm{1} x_0^\diamond  
    }+\ba}_i}
  \end{IEEEeqnarray}
  where we have used the factorization \eqref{eq:PsiFGgen}
  with $F$ obtained by the block partitioning $\Psi = \SBRbig{F \; 
  \bm{0}_{p\times1}  }$, $Z=\SBRbig{I_{p-1} \; \bm{0}_{(p-1)\times1}}$, and 
  the fact that $F^+$ is equal to the $(p-1)\times p$ lower-triangular 
  matrix of ones.  When  \eqref{eq:condint} holds, $\bm{1}x_0^\diamond\in  
  \cX^\diamond\cap \interior C$,
  Corollary~\ref{cor:xstarintC} applies,
  $\ba = \bm{0}$ (see \eqref{eq:NCzero}), and \eqref{eq:UTV} reduces to:
  \begin{IEEEeqnarray}{rCl}
    \label{eq:UTV2}
    U=\max_{1 \leq j < p}
    \absbigg{\sum_{i=1}^j \SBRbig{\nabla\cL\PARENSs{ \bm{1} x_0^\diamond 
    }}_i}.
  \end{IEEEeqnarray}
\end{subequations}

  %which can be obtained by appending a column of zeros to the right of the 
  %$(J-1) \times (J-1)$ running-sum matrix 
  %\cite[Sec.~6.4]{BoydUnpublishedBook}.  

%??? when is $U=0$?
%\\A: when $\nabla\cL(\bx^\diamond)=\bm{0}$.

  %  ??? here, we can expect this bound to differ from the ADMM one?

\end{exa}

The bounds obtained by solving (P$_0$) are often simple but restricted 
to the scenario where \eqref{eq:cond} holds.  In the following section, we 
remove assumption  \eqref{eq:cond} and develop a general numerical method 
for finding $U$ in \eqref{eq:Udef}.

\section{ADMM Algorithm for Computing $U$}
\label{sec:alg}

We focus on the nontrivial scenario where \eqref{eq:noNeed4r}
does not hold and assume $u>0$.
We also assume that an $\bx^\diamond \in \mathcal{X}^\diamond$ is 
available, which will be sufficient to obtain the $U$ in \eqref{eq:Udef}.
We use the duality of norms \cite[App.~A.1.6]{Boyd2004}:
\begin{IEEEeqnarray}{c}
  \label{eq:normduality}
  \norm{\Psi^H\bx}_1 = \max_{\|\bw\|_\infty\leq1} 
  \Realpart\PARENSs{\bw^H\Psi^H\bx}
\end{IEEEeqnarray}
to rewrite the minimization of \eqref{eq:penalizedNLL} as the following 
min-max problem (see also \eqref{eq:H}):
\begin{IEEEeqnarray}{rCl}
  \label{eq:minmax}
  \min_{\bx} \max_{\bw} \cL(\bx)+u \Realpart\PARENSs{\bw^H\Psi^{H}\bx}
  +\mathbb{I}_C(\bx)-\mathbb{I}_H(\bw).
  \IEEEeqnarraynumspace
\end{IEEEeqnarray}
Since the objective function in \eqref{eq:minmax} is convex with respect to 
$\bx$ and concave with respect to $\bw$, the optimal $(\bx,\bw)= 
\PARENSs{\bx_u,\bw_u}$ is at the saddle point of
\eqref{eq:minmax} and satisfies
\begin{subequations}
  \label{eq:TVcond}
  \begin{IEEEeqnarray}{rCl}
    \label{eq:TVcond0}
    \bm{0}&\in&\nabla\cL\PARENSs{\bx_u}
    +u\Realpart\PARENSs{\Psi \bw_u}
    +N_C\PARENSs{\bx_u}
    \\
    \label{eq:dWRTpq}
    %\bm{0}&\in& \Psi^T\bx_u - N_H\PARENSs{\bw_u}
    \bw_u&\in&G(\Psi^{H}\bx_u).
  \end{IEEEeqnarray}
\end{subequations}
%A: no need for \eqref{eq:rmw} any more.
%
%
%\begin{IEEEeqnarray}{rCl}
%  \label{eq:A}
%  \mathcal{A} &\df& \CBRbig{i\in\{1,2,\ldots,p'\} \mid 
%  \SBRs{\Psi^T\bx^\diamond}_i\neq0}.
%\end{IEEEeqnarray}
%This result follows by using (see \eqref{eq:H})
%\begin{IEEEeqnarray}{c} [N_H(\bm{\mathrm{w})}]_i=\left\{
%    \begin{IEEEeqnarraybox}[][c]{r?l}
%      \IEEEstrut
%      (-\infty,0\rbrack, & \mathrm{w}_i=-1,\\
%      \lbrack0,+\infty), & \mathrm{w}_i=1,\\
%      \{0\},       &  \mathrm{w}_i \in (-1,1)
%      \IEEEstrut
%    \end{IEEEeqnarraybox}
%  \right..
%\end{IEEEeqnarray}
Now, select $U$ as the smallest $u$ for which 
\eqref{eq:TVcond0}--\eqref{eq:dWRTpq} hold with $\bx_u=\bx^\diamond$:
\begin{IEEEeqnarray}{c}
  \label{eq:Uas1overvstarTV}
  U=  \frac{1}{v^\diamond}
 {\norm{\nabla\cL\PARENSs{\bx^\diamond}}_2}
 \end{IEEEeqnarray}
where $\PARENSs{v^\diamond, \bw^\diamond, \bt^\diamond}$ is the solution to 
the following constrained linear programming problem:
\begin{IEEEeqnarray}{s'l+l+x*}
  \IEEEyesnumber
  \label{eq:problem}
  \IEEEyessubnumber
%\begin{aligned}
   (P$_1$): & \underset{v,\bw,\bt}{\text{minimize}}
  &
  -v
  +\mathbb{I}_{G(\Psi^{H}\bx^\diamond)}(\bw)
  +\mathbb{I}_{N_C\PARENSs{\bx^\diamond}}(\bt) \IEEEeqnarraynumspace
  \\
  \IEEEyessubnumber
  & \text{subject to}
  & \kern20pt
  \label{eq:sumZero}
  v\bg+\Realpart\PARENSs{\Psi\bw}+ \bt=\bm{0}
 %\\
 %\IEEEyessubnumber
 %\label{eq:rmw2}
 %& \kern10pt
 %\bw_{{\mathcal{A}}}=
 %\bm{\rm{w}}_{{\mathcal{A}}}
\end{IEEEeqnarray}
obtained from \eqref{eq:TVcond0}--\eqref{eq:dWRTpq}
with $\bx_u$ and $\bw_u$ replaced by $\bx^\diamond$ and $\bw$.  Here, 
\begin{IEEEeqnarray}{c}
  \bg \df {\nabla\cL\PARENSs{\bx^\diamond}} \big/ 
  {\norm{\nabla\cL\PARENSs{\bx^\diamond}}_2}
  \label{eq:bg}
\end{IEEEeqnarray}
is the normalized gradient (for numerical stability) of the \gls{NLL} at 
$\bx^\diamond$;
$\nabla\cL\PARENSs{\bx^\diamond} \neq \bm{0}$ because \eqref{eq:noNeed4r} 
does not hold.  Due to \eqref{eq:bxStarCond}, $v=0$ is a feasible point 
that satisfies the constraints \eqref{eq:sumZero}, which implies that 
$v^\diamond\geq0$.  When \eqref{eq:Uinftycond} holds, $v$ has to be zero, 
implying $U=+\infty$.

To solve (P$_1$) and find ${v^\diamond}$, we apply an iterative algorithm 
based on \gls{ADMM} \cite{Boyd2011ADMM,HongLuo2013ADMM}
\begin{subequations}
\label{eq:admm}
\begin{IEEEeqnarray}{rCl}
  \label{eq:quadraticBox}
  \bw^{(i+1)} &=&\arg \kern-10pt \min_{
    \bw\in G\PARENSs{\Psi^{H}\bx^\diamond}
  }  \norm{
    v^{(i)}\bg+
    \Realpart\PARENSs{\Psi\bw} +\bt^{(i)} +{\bz^{(i)}}
  }_2^2
  \IEEEeqnarraynumspace
  \\
  v^{(i+1)}&=&\rho
  -\bg^T\SBRbig{\Realpart\PARENSs{\Psi\bw^{(i+1)}}+\bt^{(i)}+{\bz^{(i)}}}
  \IEEEeqnarraynumspace
  \\
  \label{eq:t}
  \bt^{(i+1)}&=&P_{N_C\PARENSs{\bx^\diamond}}\PARENSbig{
    -{v^{(i+1)}\bg-\Realpart\PARENSs{\Psi\bw^{(i+1)}}-{\bz^{(i)}}}
  }
  \\
  \bz^{(i+1)}&=&\bz^{(i)}+{
    \Realpart\PARENSs{\Psi\bw^{(i+1)}}
    +v^{(i+1)}\bg
    +\bt^{(i+1)}
  }
\end{IEEEeqnarray}
\end{subequations}
where $\rho>0$ is a tuning parameter for the \gls{ADMM} iteration and we 
solve \eqref{eq:quadraticBox} using the
 Broyden-Fletcher-Goldfarb-Shanno
optimization algorithm  with box constraints
\cite{Byrd1995LBFGS} and \gls{PNPG} algorithm  \cite{PNPGtechreport} for 
real and complex $\Psi$, respectively.  We initialize the iteration 
\eqref{eq:admm} with $v^{(0)}=1$, $\bt^{(0)}=\bm{0}$,  $\bz^{(0)}=\bm{0}$, 
and $\rho=1$, where $\rho$ is adaptively adjusted
thereafter using the scheme in \cite[Sec.~3.4.1]{Boyd2011ADMM}.

%??? We initialize $\rho$ by 1 and adaptively adjust its value thereafter 
%using the scheme in \cite[Sec.~3.4.1]{Boyd2011ADMM}.
%A: This is not trivial when it comes to have three variables to alternate.  
%Better to just use 1.

In special cases, \eqref{eq:admm} simplifies.   If \eqref{eq:cond} holds, 
then $\Psi^{H} \bx^\diamond = \bm{0}$ and the constraint in 
\eqref{eq:quadraticBox} simplifies to $\norm{\bw}_\infty\leq1$; see 
\eqref{eq:H}.  If $\Realpart(\Psi\Psi^H)=cI, \, c>0$, and 
$\Psi\in\mathamsbb{R}^{p\times p}$ or $\Psi\in\mathamsbb{C}^{p\times p/2}$,
%
%{???A: when $\Realpart(\Psi\Psi^H)=cI$, we have
%  \begin{equation}
%    v^{(i)}\bg+
%    \Realpart\PARENSs{\Psi\bw} +\bt^{(i)} +{\bz^{(i)}}
%    \\
%    =
%    \Realpart\PARENSbigg{
%      \Psi\SBRbig{
%        \bw+\frac{1}{c}
%        \Psi^H\PARENSs{v^{(i)}\bg+\bt^{(i)} +{\bz^{(i)}}}
%      }
%    }
%  \end{equation}
%  Now, when $\Psi$ is real, it it clear that we want $\Psi^T\Psi$ be 
%  proportinal to $I$, which implies that $\Psi$ is square.
%  When $\Psi$ is complex, 
%  $\Realpart(\Psi\Psi^H)=\underline{\Psi}\underline{\Psi}^T$ and we want 
%  $\underline{\Psi}^T\underline{\Psi}$ also be proportional to $I$, which 
%  means $\underline{\Psi}$ has to be square.  Is there anyway to unite the 
%  two?
%}
%
\eqref{eq:quadraticBox} has the following analytical solution:
\begin{IEEEeqnarray}{c}
  \label{eq:wanalytical}
  \bw^{(i+1)} =\projpBig{G\PARENSs{\Psi^{H}\bx^\diamond}}
  {-\frac{1}{c}\Psi^H\PARENSbig{v^{(i)}\bg+\bt^{(i)}+{\bz^{(i)}}}}.
\end{IEEEeqnarray}
When \eqref{eq:condint} holds, \eqref{eq:t} reduces to $\bt^{(i)}=\bm{0}$ 
for all $i$, thanks to \eqref{eq:NCzero}.

When $\Psi$ is real, the constraints imposed by 
$\mathbb{I}_{G(\Psi^H\bx^\diamond)}(\bw)$ become linear and (P$_1$)  
becomes a linear programming problem with linear constraints.

\begin{table*}
  \centering
  \begin{tabular}{r r r r r r r r r}
       &\multicolumn{2}{c}{$C=\mathamsbb{R}_+^p$, DWT}
    &\multicolumn{2}{c}{$C=\mathamsbb{R}^p$, DWT}
    &\multicolumn{2}{c}{$C=\mathamsbb{R}_+^p$, TV}
    &\multicolumn{2}{c}{$C=\mathamsbb{R}^p$, TV}
    \\
      SNR/dB
  &{theoretical}&{empirical}
    &{theoretical}&{empirical}
    &{theoretical}&{empirical}
    &{theoretical}&{empirical}
    \\
    \cmidrule(l){1-1}
    \cmidrule(l){2-3}
    \cmidrule(l){4-5}
    \cmidrule(l){6-7}
    \cmidrule(l){8-9}
    \num{ 30}& 8.87& 8.87&  9.43&  9.43& 101.55& 
    101.54&\multicolumn{2}{c}{\multirow{7}{*}{\parbox[c]{60pt}{
          \centering
          same as\\
          $C=\mathamsbb{R}_+^p$, TV
    }}}\\
    \num{ 20}& 8.91& 8.91&  9.47&  9.47& 100.21& 100.21&       &       \\
    \num{ 10}& 9.03& 9.03&  9.59&  9.59&  96.47&  96.47&       &       \\
    \num{  0}& 9.43& 9.43&  9.98&  9.98&  87.49&  87.49&       &       \\
    \num{-10}&11.88&11.89& 14.03& 14.02& 152.07& 152.07&       &       \\
    \num{-20}&27.77&27.78& 43.28& 43.28& 361.56& 361.56&       &       \\
    \num{-30}&88.78&88.82&139.67&139.66&1024.04&1024.04&       &       \\
    \num{-30}&77.29&77.31&123.91&123.90& 
    683.43%\makebox[0cm][l]{$^\dagger$}
                                                &683.43& 909.50& 909.48\\
 %   \cmidrule(l){6-6}
 %   \multicolumn{5}{c}{ }&\multicolumn{4}{l}{
 %     \makebox[0cm][l]{$^\dagger$ $x_0<0$ defined in \eqref{eq:x0}}
 %   }
  \end{tabular}
  \caption{Theoretical and empirical bounds $U$ for the linear Gaussian 
  model.}
  \label{table:slBound}
\end{table*}

\sisetup{retain-zero-exponent=true}
\begin{table*}
  \centering
  \begin{tabular}{l l l l l l l}
    &\multicolumn{2}{c}{DWT}
    &\multicolumn{2}{c}{Anisotropic TV}
    &\multicolumn{2}{c}{Isotropic TV}
    \\
    $\bm{1}^T\Phi\bx_\text{true}$
    &{theoretical}&{empirical}
    &{theoretical}&{empirical}
    &{theoretical}&{empirical}
    \\
    \cmidrule(l){1-1}
    \cmidrule(l){2-3}
    \cmidrule(l){4-5}
    \cmidrule(l){6-7}
    %\num{e0}&\num{4.769e-2}&\num{4.775e-2}&\num{1.828e-2}&\num{1.821e-2}&\num{2.035e-2}&\num{2.028e-2}\\
    \num{e1}&\num{9.660e-1}&\num{9.662e-1}&\num{7.550e-2}&\num{7.544e-2}&\num{7.971e-2}&\num{7.937e-2}\\
    %\num{e2}&\num{1.244e+1}&\num{1.244e+1}&\num{4.162e-1}&\num{4.161e-1}&\num{4.778e-1}&\num{4.765e-1}\\
    \num{e3}&\num{1.155e+2}&\num{1.156e+2}&\num{4.154e+0}&\num{4.153e+0}&\num{4.888e+0}&\num{4.877e+0}\\
    %\num{e4}&\num{1.164e+3}&\num{1.164e+3}&\num{3.994e+1}&\num{3.993e+1}&\num{4.719e+1}&\num{4.709e+1}\\
    \num{e5}&\num{1.153e+4}&\num{1.153e+4}&\num{3.951e+2}&\num{3.950e+2}&\num{4.666e+2}&\num{4.656e+2}\\
    %\num{e6}&\num{1.144e+5}&\num{1.144e+5}&\num{3.949e+3}&\num{3.948e+3}&\num{4.665e+3}&\num{4.654e+3}\\
    \num{e7}&\num{1.145e+6}&\num{1.145e+6}&\num{3.947e+4}&\num{3.946e+4}&\num{4.661e+4}&\num{4.651e+4}\\
    %\num{e8}&\num{1.145e+7}&\num{1.145e+7}&\num{3.988e+5}&\num{3.987e+5}&\num{4.709e+5}&\num{4.696e+5}\\
    \num{e9}&\num{1.153e+8}&\num{1.154e+8}&\num{3.950e+6}&\num{3.949e+6}&\num{4.665e+6}&\num{4.654e+6}
  \end{tabular}
  \caption{Theoretical and empirical bounds $U$ for the PET example.  }
  \label{table:petBound}
\end{table*}

\section{Numerical Examples}
\label{sec:NumEx}

Matlab implementations of the presented examples are available at 
\url{https://github.com/isucsp/imgRecSrc/uBoundEx}.
%In the following, whenever we use \gls{PNPG} method \cite{PNPGtechreport}, 
%we keep all the parameters to their default unless specified otherwise.
In all numerical examples,
the empirical upper bounds $U$ were obtained by a grid search over $u$ with 
$\mathcal{X}_u =\{\bx_u\}$ obtained using the \gls{PNPG} method 
\cite{PNPGtechreport}.

\subsection{Signal reconstruction for Gaussian linear model}
%\label{sec:skyline}

We adopt the linear measurement model with white Gaussian noise and
scaled \gls{NLL}  $\cL(\bx) = 0.5 \|\by-\Phi\bx\|_2^2$,
where the elements of the sensing matrix $\Phi \in \mathamsbb{R}^{N \times 
p}$ are \gls{iid} and drawn from the uniform distribution on a unit sphere.  
We reconstruct the nonnegative ``skyline'' signal $\bx_\text{true} \in 
\mathamsbb{R}^{1024 \times 1}$  in 
\cite[Sec.~\ref{report-sec:linear1dex}]{PNPGtechreport} from noisy linear 
measurements $\by$ using the \gls{DWT} and
1D \gls{TV} regularizations, where
the \gls{DWT} matrix $\Psi$ is orthogonal  ($\Psi\Psi^T = \Psi^T \Psi = 
I$),
constructed using the Daubechies-4 wavelet with three decomposition levels.
Define the \gls{SNR} as
\begin{equation}
  \SNR \, (\si{\text{\decibel}}) = 
  10\log_{10}{\frac{\|\Phi\bx_\text{true}\|_2^2}{N\sigma^2}}
  \label{eq:defSNR}
\end{equation}
where $\sigma^2$ is the variance of the Gaussian noise added to $\Phi 
\bx_\text{true}$ to create the noisy measurement vector $\by$.

For $C=\mathamsbb{R}_+^p$ and $C=\mathamsbb{R}^p$ with \gls{DWT} 
regularization, $\mathcal{X}^\diamond=\{\bm{0}\}$ and 
Example~\ref{ex:nonnegsig} applies and yields the upper bounds 
\eqref{eq:UCnonnegZsquare} 
%
%(which becomes $U = 
%\min_{\ba\succeq0}\norm{\Psi^T\PARENSs{\Phi^T\by+\ba}}_\infty$, computed 
%using CVX \cite{cvx}) 
%
and \eqref{eq:UinvertiblerealPsizerointC}, respectively.

For \gls{TV} regularization, we apply the result in Example~\ref{ex:1DTV}.  
For $C=\mathamsbb{R}^p$ and $C=\mathamsbb{R}_+^p$, we have   
$\mathcal{X}^\diamond=\{  \bm{1} x_0\}$ and $\mathcal{X}^\diamond=\{  
\bm{1} \max(x_0,0) \}$, respectively, where
\begin{IEEEeqnarray}{rCl}
  \label{eq:x0}
  x_0 \df  \arg \min_{x \in \mathamsbb{R}}  \cL(\bm{1} x)  = 
  \bm{1}^T\Phi^T\by / \norm{\Phi\bm{1}}^2_2.
\end{IEEEeqnarray}
If $\bm{1} x_0 \in \interior C$, which holds when $C=\mathamsbb{R}^p$ or 
when $C=\mathamsbb{R}_+^p$ and $x_0 > 0$, then the bound $U$ is given by  
\eqref{eq:UTV2}.
 For  $C=\mathamsbb{R}_+^p$ and if $x_0 \leq 0$, then
$\mathcal{X}^\diamond=\{  \bm{0} \}$ and \eqref{eq:UTV} applies.   In this 
case, $U=0$ if $[\nabla\cL(\bm{0})]_i \geq 0$ for $i=1,\dotsc,p-1$, which 
occurs only when $[\nabla\cL(\bm{0})]_i=0$ for all $i$.

%\begin{IEEEeqnarray}{c}
%  \label{eq:UTVex}
%  U = \min_{ \ba \preceq \bm{0}  } \max_{1 \leq j < p}
%  \absbigg{\sum_{i=1}^j \SBRbig{\nabla\cL\PARENSs{\bm{0}}+\ba}_i}  
%\end{IEEEeqnarray}

Table~\ref{table:slBound} shows the theoretical and empirical bounds for 
\gls{DWT} and \gls{TV} regularizations and $C=\mathamsbb{R}_+^p$ and
$C=\mathamsbb{R}^p$; we decrease the \gls{SNR} from 
\SIrange{30}{-30}{\decibel} with independent noise realizations for 
different \glspl{SNR}.  The theoretical bounds in 
Sections~\ref{sec:theorem}
and \ref{sec:alg} coincide.
%
%\eqref{eq:UCnonnegZsquare}, \eqref{eq:U0linGauss}, \eqref{eq:UTV}, and 
%\eqref{eq:UTV2} and corresponding bounds \eqref{eq:Uas1overvstarTV} 
%
%(obtained using the \gls{ADMM}-type iteration \eqref{eq:admm} with 
%\eqref{eq:wanalytical}, because $\Psi$ is orthogonal) 
%
%coincide.
For \gls{DWT} regularization, $\cX^\diamond$ is the same for both convex 
sets $C$ and thus the upper bound $U$ for $C=\mathamsbb{R}_+^p$ is always 
smaller than its counterpart for $C=\mathamsbb{R}^p$, thanks to being 
optimized over variable $\ba$ in \eqref{eq:UCnonnegZsquare}.
For \gls{TV} regularization, when $x_0>0$, the upper bounds $U$ coincide 
for both $C$ because, in this case,
$\cX^\diamond$ is the same for both $C$ and $\cX^\diamond\in\interior C$.  
In the last row of Table~\ref{table:slBound} we show the case where 
$x_0\leq0$; then, $\cX^\diamond$ differs for the two convex sets $C$,
and the upper bound $U$ for $C=\mathamsbb{R}_+^p$ is smaller than its 
counterpart for $C=\mathamsbb{R}^p$, thanks to
being optimized over variable
 $\ba$ in \eqref{eq:UTV}: compare \eqref{eq:UTV} with \eqref{eq:UTV2}.

%Comment on the fact that TV bound is not monotonic
%with SNR. Any intuition about this?
%A: maybe the following. But I am not sure. It must settle somewhere as SNR 
%goes up.  I guess it is something similar to dithering. I will run more to 
%see. and have another implementation of the noises.\\
%When the measurements becomes highly accurate, the information provided via 
%\gls{NLL} grows, which counters more the regularization that we enforce on 
%$\bx$ wishing it to be constant.

%See also the second column of Table~\ref{table:slBound}, which is obtained 
%by \eqref{eq:admm}.  The same solution is achieved by using CVX \cite{cvx} 
%directly from \eqref{eq:UlinGauss}.
%
%We show the empirical solution in the third column of 
%Table~\ref{table:slBound}, which is obtained via the bisection method by 
%running the reconstruction algorithm \cite{PNPGtechreport} initialized by 
%zero and judging whether the solutions alters.  Due to the numerical 
%problem caused by inaccuracy of the inner iteration, the empirical $U$s are 
%slightly smaller than the theoretical ones.

%\renewcommand{\arraystretch}{1.2}

%Select $C=\mathamsbb{R}^p$: applying \eqref{eq:U0linGauss} to obtain the 
%fourth column of Table~\ref{table:slBound}, which is also achievable by 
%\eqref{eq:admm}.  The fifth column in Table~\ref{table:slBound} by 
%bisection method is exactly the same as the theoretical bound because of 
%the exact proximal step.
%
%

\subsection{PET image reconstruction from Poisson measurements}

Consider \gls{PET} reconstruction of the $128\times128$ concentration map 
$\bx_\text{true}$ in \cite[Fig.~\ref{report-fig:pet}]{PNPGtechreport}, 
which represents simulated radiotracer activity in a human chest, from 
independent noisy Poisson-distributed measurements $\by=(y_n)$ with means 
$[\Phi\bx_\text{true}+\bb]_n$.  The choices of parameters in the \gls{PET} 
system setup and concentration map $\bx_\text{true}$ have been taken from 
the \gls{IRT} \cite[\texttt{emission/em\_test\_setup.m}]{irt}.  Here,
\begin{subequations}
\begin{IEEEeqnarray}{c}
  \cL(\bx)=\bm{1}^T \PARENSs{\Phi\bx+\bb-\by}
  + \sum_{n,y_n\neq0}{y_n}\ln\frac{y_n}{\SBRs{\Phi\bx+\bb}_n}
  \label{eq:poissonl}
  \IEEEeqnarraynumspace
\end{IEEEeqnarray}
and
\begin{IEEEeqnarray}{c}
  \Phi= w \diag\bigl(\exp_\circ(-S\bkappa+\bc)\bigr)S
  \in \mathamsbb{R}_+^{N \times p}
  \label{eq:petPhi}
\end{IEEEeqnarray}
\end{subequations}
is the known sensing matrix; $\bkappa$ is the density map
needed to model the attenuation of the gamma rays 
\cite{OllingerFessler1997PET}; $\bb=(b_i)$ is the known intercept term 
accounting for background radiation, scattering effect, and accidental 
coincidence;\footnote{The elements of the intercept term have been set to a 
  constant equal to \SI{10}{\percent} of the sample mean of $\Phi 
\bx_\text{true}$: $\bb=[{\bm{1}^T\Phi \bx_\text{true} }/(10N)]\bm{1}$.}
$\bc$ is a known vector that models the detector efficiency variation; and 
$w>0$ is a known scaling constant, which we use to control the expected 
total number of detected  photons due to electron-positron annihilation, 
$\bm{1}^T\Exp(\by-\bb) = \bm{1}^T\Phi\bx_\text{true}$, an \gls{SNR} 
measure.  We collect the photons from \num{90} equally spaced directions 
over  \SI{180}{\degree}, with \num{128} radial samples at each direction.  
Here,   we adopt the parallel strip-integral matrix $S$ 
\cite[Ch.~25.2]{FesslerUnpublishedBook} and use its implementation in the 
\gls{IRT} \cite{irt}.

We now consider the nonnegative convex set $C=\mathamsbb{R}_+^p$,
which ensures that \eqref{eq:Ccond} holds, and 2D isotropic and anisotropic 
\gls{TV} and \gls{DWT} regularizations, where the 2D \gls{DWT} matrix 
$\Psi$ is constructed using the Daubechies-6 wavelet with six decomposition 
levels.

For \gls{TV} regularizations, $\mathcal{X}^\diamond=\{  \bm{1} \max(0,x_0) 
\}$, where $x_0 = \arg \min_{x \in \mathamsbb{R}} \cL(\bm{1} x)$, computed 
using the bisection method that finds the zero of
${\partial\cL(\bm{1}x)}/{\partial x}$,
which is an increasing function of $x\in\mathamsbb{R}_+$.  Here, no search 
for $x_0$ is needed when $\eval[0]{\partial\cL(\bm{1}x) / \partial x}_{x=0}  
> 0$, because in this case $x_0<0$.

%For \gls{TV} regularizations, $\mathcal{X}^\diamond=\{  \bm{1} \max(0,x_0) 
%\}$, where $x_0 = \arg \min_{x \in \mathamsbb{R}} \cL(\bm{1} x)$, computed 
%using the bisection method \cite[Sec.~4.2.4]{Thisted} that finds the zero 
%of
%\begin{IEEEeqnarray}{rCl}
%  \label{eq:partialNLL}
%  \frac{\partial\cL(\bm{1}x)}{\partial x}
%  =
%  \bm{1}^T\Phi\bm{1}-\by^T\diag^{-1}\PARENSs{\Phi\bm{1}x+\bb}\Phi\bm{1}
%\end{IEEEeqnarray}
%which is an increasing function of $x\in\mathamsbb{R}_+$.  Here, no search 
%for $x_0$ is needed when $\eval[0]{\partial\cL(\bm{1}x) / \partial x}_{x=0}  
%> 0$, because in this case $x_0<0$. 

We computed the theoretical bounds using the \gls{ADMM}-type algorithm in 
Section~\ref{sec:alg}.

Table~\ref{table:petBound} shows the theoretical and empirical bounds for 
\gls{DWT} and \gls{TV} regularizations and the \gls{SNR} 
$\bm{1}^T\Phi\bx_\text{true}$
varying from \num{e1} to \num{e9},  with independent measurement 
realizations for different \glspl{SNR}.

%??? In all examples that we ran,  $x_0>0$. ??? or maybe not, you can try 
%more.
%
%??? I think the table has too many rows, you can put $0.1, 10, 1000, 10^5 
%10^7 10^9$. Maybe try 0.01? At least these small values break the 10x 
%pattern.  Maybe you will hit $x_0<0$.
%\\A: smaller than 1 does not make sense, which is equivalent to set all 
%the measurement to zero.

%\\A: tested a few, when summation of all element of $y=0$, $x_0=0$; even 
%when sum(y)=1, $x_0>0$.
%Note that the first row that I show in the table has sum(y)=1, in which 
%case, I don't think this make much sense.  Better to keep rows from the 
%3rd.

Denote the isotropic and anisotropic 2D \gls{TV} bounds by $U_\text{iso}$ 
and $U_\text{ani}$, respectively. Then, it is easy to show that when 
\eqref{eq:cond} holds,
$U_\text{ani}\leq U_\text{iso}\leq\sqrt{2}U_\text{ani}$, which follows by 
using the inequalities 
$\sqrt{2}\sqrt{a^2+b^2}\geq\abs{a}+\abs{b}\geq\sqrt{a^2+b^2}$ and is 
confirmed in Table~\ref{table:petBound}.

%how is this called? A: I don't know a exact name, but it can be easily 
%obtained from Inequality of arithmetic and geometric means.

%\begin{IEEEproof}
%  Thanks to \eqref{eq:cond}, isotropic and anisotropic \gls{TV} share the 
%  same $\cX^\diamond$.  Suppose $\bm{1}x_0^\diamond\in\cX^\diamond$, we 
%  have
%  \begin{IEEEeqnarray}{rCl}
%    \cL(\bm{1}x_0^\diamond)\leq f_{U_\text{ani}}(\bx)
%  \end{IEEEeqnarray}
%  for any $\bx\in\dom\cL$.  Use the fact
%  $\sqrt{2}\sqrt{a^2+b^2}\geq\abs{a}+\abs{b}$ to get
%  $r_\text{ani}(\bx)\leq \sqrt{2}r_\text{iso}(\bx)$.  Therefore, 
%  \begin{IEEEeqnarray}{rCl}
%    \cL(\bm{1}x_0^\diamond)\leq f_{\sqrt{2}U_\text{ani}}(\bx)
%  \end{IEEEeqnarray}
%  for all $\bx$ and $U_\text{iso}\leq \sqrt{2}U_\text{ani}$.
%\end{IEEEproof}

%\begin{figure}
%  \centering
%  \includegraphics{bound4U}
%  \caption{Comparison between $U$ and the empirical bound for anisotropic 
%    \gls{TV}.}
%  \label{fig:bound4U}
%\end{figure}

\section{Concluding Remarks}
\label{sec:conclusion}

Future work will include obtaining simple expressions for upper bounds $U$ 
for isotropic 2D \gls{TV} regularization, based on Theorem~\ref{thm:U}.

\printbibliography
%\balance
\end{document}